# Structural Phase Transformations in Iron-Chalcogen under High Pressures

Andrew K. Stemshorn and Yogesh K. Vohra

Department of Physics, University of Alabama at Birmingham (UAB)

Birmingham, AL 35294, USA

Phillip M. Wu

Department of Physics, Duke University

Durham, NC 27708, USA

F.C. Hsu, Y.L. Huang, M.K. Wu and K.W. Yeh

Institute of Physics, Academia Sinica – Nankang

Taipei, TAIWAN

ABSTRACT

We report high pressure structural phase transformation sequence in a layered Iron-based superconducting compound $FeSe_{0.3}Te_{0.7}$ to 31 GPa at room temperature. The ambient pressure PbO type tetragonal phase (Space Group – P4/nmm) transforms to a monoclinic phase (Space group - $P2_1/m$) at a pressure of 7.3 ± 0.9 GPa. This monoclinic phase is similar to the one observed below 100 K at ambient pressure. On further increase of pressure above 12 GPa, a transformation to an amorphous phase is observed that is completed by 20 GPa. The amorphous phase is found to be stable to the highest pressure of 31 GPa. This structural sequence of tetragonal → monoclinic → amorphous phase transformation appears to be a common feature of iron-based layered superconductors under compression. The pressure induced structural phase transformations are likely to play a key role in the superconductivity in these materials at high pressures.

...

The superconductivity in the PbO-type structure β-FeSe was discovered recently at 8 K in samples prepared with Se deficiency [1]. The occurrence of superconductivity in this simple layered FeSe compound with edge-sharing $FeSe_4$ tetrahedron has created tremendous interest in the effects of chemical substitution and high pressure on this material system [2]. In particular, Tellurium (Te) substitution has been studied [3, 4] and a maximum superconducting transition temperature of 15.2 K was measured for $FeSe_{0.5}Te_{0.5}$. Additionally, pressure induced enhancement of superconducting transition temperature of FeSeTe system has been reported [4] that warrant further high pressure structural investigations. A recent high pressure study [5] on the $FeSe_{0.5}Te_{0.5}$ compound has reported a pressure induced tetragonal to amorphous phase transformation at a pressure of 11.5 GPa. This pressure induced amorphization in $FeSe_{0.5}Te_{0.5}$ compound was found to be reversible on decreasing pressure and a tetragonal phase was recovered during decompression. There is interest in exploring whether this pressure induced amorphization phenomenon is observed at other compositions in $FeSe_xTe_{(1-x)}$ compounds and to establish the correlation between the crystallographic phases and the superconducting transition temperature in this material system.

Powder materials of Fe (3N purity), Se (3N purity), and Te (5N purity) with appropriate stoichiometry ($FeSe_{0.3}Te_{0.7}$) were mixed in a ball mill for a duration of one hour. The well mixed powders were cold-pressed into discs under 400 kg/cm$^2$ uniaxial pressure, and then sealed in an evacuated quartz tube with pressure less than $10^{-4}$ torr and heat treated at $600^0$-C for 20 hours. The reacted bulk sample was reground into fine powders repressed, sealed, and subsequently sintered at $650^0$-C for 20 hours. The sample for high pressure studies was cut from the sintered pellet and loaded in the diamond anvil cell. The $FeSe_{0.3}Te_{0.7}$ compound belongs to the PbO type structure (Space Group P4/nmm), tetragonal with 4 atoms/cell. Fe atoms occupy the 2a positions (3/4, 1/4, 0) and (1/4, 3/4, 0) and Se and Te atoms at random occupy 2c positions (1/4, 1/4, z), and (3/4, 3/4, -z). The parameter z has been determined to be 0.2715 based on earlier work on $Fe_{1+y}Se_xTe_{1-x}$ compounds [6].

The high pressure x-ray diffraction experiments were carried out at the beam-line 16-BM-D, HPCAT, Advanced Photon Source, Argonne National Laboratory. An angle dispersive technique with an image-plate area detector was employed using a x-ray wavelength λ = 0.3875

Å. The high pressure sample to image plate detector distance was set to 366.3 mm. The polycrystalline powder of $FeSe_{0.3}Te_{0.7}$ compound was placed in an 80 micron diameter sample hole in a spring steel gasket in a diamond anvil cell device employing diamond anvils with culet size of 300 microns. A copper foil was also placed in the sample chamber to serve as an internal x-ray pressure standard. We did not employ any pressure transmitting medium to avoid interference from weak extraneous peaks in our crystal structure determinations by x-ray diffraction methods.

In order to obtain the most accurate pressure from our internal copper pressure marker, Birch-Murnaghan equation of state (EoS) [7] shown in equation (1) was fitted to the available data on copper and details on the fit are provided in reference [8].

$$P = 3B_0 f_E (1+2f_E)^{5/2} \left\{ 1 + \frac{3}{2}(B'-4)f_E \right\} \qquad (1)$$

Where $B_o$ is the bulk modulus, B' is the first derivative of bulk modulus at ambient pressure, and $V_0$ is the ambient pressure volume. Our measured value of ambient pressure volume for copper is 11.802 $Å^3$/atom. The parameter $f_E$ is related to volume compression and is described below.

$$f_E = \frac{\left[ \left( \frac{V_o}{V} \right)^{2/3} - 1 \right]}{2}$$

The volume of copper pressure marker was measured at high pressures and the pressure was calculated from the known equation of state of copper fitted to the Birch-Murnaghan equation (1) with Bulk Modulus ($B_0$) = 121.6 GPa and its pressure derivative B' = 5.583.

Fig.1 shows the integrated diffraction profiles for $FeSe_{0.3}Te_{0.7}$ sample along with copper pressure marker at various pressures recorded with a x-ray wavelength λ = 0.3875 Å. The copper pressure marker is in the face centered cubic phase in the entire pressure range as evidenced by the Cu (111), (200), (220), (311), and (222) diffraction peaks. A least squared fit to the cubic phase of copper results in a lattice parameter at high pressure which in turn yields atomic volume and a measured pressure value using the equation of state discussed above. The $FeSe_{0.3}Te_{0.7}$ sample is in the tetragonal P4/nmm phase at a pressure of 0.53 GPa with lattice parameters a = 3.807 Å and c = 6.114 Å and c/a = 1.606 (Fig. 1 (a)). All the diffraction peaks in Fig. 1 (a) from the $FeSe_{0.3}Te_{0.7}$ sample can be assigned to a tetragonal phase. As pressure is increased to 7.3 ± 0.9 GPa, a phase transformation is observed as evidenced by the splitting of a major (101) peak of the tetragonal phase. This splitting is clearly visible is the x-ray diffraction spectrum at 12.2 GPa and this new phase can be indexed to a monoclinic $P2_1/m$ phase as shown in Fig. 1 (b). The measured lattice parameters for the monoclinic phase at 12.2 GPa are a = 3.656 Å, b = 3.589 Å, and c = 5.850 Å, and angle β = 91.3 degrees. The monoclinic $P2_1/m$ phase has 4 atoms/cell with Fe atoms occupying the 2e positions (3/4, 1/4, $z_1$) and (1/4, 3/4, -$z_1$) with $z_1$ = 0.0035 and Se (occupancy = 0.3) and Te (occupancy = 0.7) atoms occupy 2e positions (1/4, 1/4, z), and (3/4, 3/4, -$z_2$) with $z_2$ = 0.2798. The parameter $z_1$ and $z_2$ has been determined earlier based on neutron diffraction studies on $Fe_{1+y}Se_xTe_{1-x}$ compounds [6]. The monoclinic $P2_1/m$ phase can be considered as a slight distortion of the P4/nmm tetragonal phase that results in *b* > *a* and angle β ≠ 90 degrees.

In addition to the monoclinic phase at 12.2 GPa, a weak broad diffraction peak marked by an asterisk is also observed. This broad diffraction peak is characteristic of an amorphous phase and grows in intensity with increasing pressure beyond 12.2 GPa with the transformation to amorphous phase completing at 20 GPa. The diffraction pattern at the highest pressure of 31 GPa in the amorphous phase is shown in Fig. 1 (c) and the three broad peaks attributed to the amorphous phase are marked by an asterisk. The strongest amorphous band observed at 31 GPa corresponds to an interplanar spacing of 2.666 Å. On decompression from the highest pressure of 31 GPa, a considerable hysteresis is observed in the phase transformations, and amorphous phase is found to be stable to pressure as low as 2.49 GPa. A representative spectrum for the amorphous phase during decompression at 5.45 GPa is shown in Fig 1 (d). On reducing pressure to 1 GPa during decompression, tetragonal P4/nmm phase is recovered and no amorphous phase is observed at ambient condition after a complete release of pressure.

Fig. 2 clearly demonstrates the gradual distortion of the tetragonal to monoclinic symmetry where axial ratio (a/b) is plotted as a function of pressure. The (a/b) has the value of 1.0 in the tetragonal phase up to 6.4 GPa and increases with increasing pressure in the monoclinic phase. The maximum (b/a) value of 1.05 or 5 % distortion is observed at a pressure of 20 GPa.

Fig. 3 shows the interplanar spacing for the amorphous phase during both compression and decompression cycles. The compression data covers the pressure range between 8.8 GPa to 31 GPa while the decompression data covers the range 1 GPa to 22.5 GPa. Fig. 2 clearly demonstrates that the amorphous phase is stable over a large pressure range and there is considerable hysteresis associated with this phase transformation. Also, during decompression, the amorphous is observed to directly transform to the tetragonal phase without going through the intermediate monoclinic phase.

Fig. 4 shows the measured equation of state (P-V data) for the superconducting compound $FeSe_{0.3}Te_{0.7}$ at room temperature. The tetragonal and monoclinic phases have been included in the data to 20 GPa. The $FeSe_{0.3}Te_{0.7}$ shows a volume compression of 20 % ($V/V_0$ = 0.8) at a pressure of 20 GPa. The measured P-V data for $FeSe_{0.3}Te_{0.7}$ sample was fitted to a Birch Murnaghan equation (1) and yielded a value of Bulk Modulus ($B_0$) = 38.6 ± 2 GPa and a pressure derivative of Bulk Modulus ($B^{'}$) = 8.1 ± 0.6 with ambient pressure volume ($V_0$) = 22.38 Å$^3$/atom.

The pressure induced structural sequences are important predictors of the high pressure superconductivity phenomenon in these materials. The crystallographic distortions and disorder reported in $FeSe_xTe_{(1-x)}$ compounds are expected to modify the electronic structure and electron-phonon coupling parameters that are so critical to the high temperature superconductivity in these materials even though the exact mechanisms remain a topic of considerable debate. The progression of phase transformations is now beginning to emerge from the high pressure studies on $FeSe_xTe_{(1-x)}$ compounds. The pressure induced amorphization phenomenon has been reported in various materials systems like ice, quartz, silicates and other minerals [9], however, there occurrence in Iron-based layered compounds has only been recently documented [6]. There appears to be a clear trend that amorphization pressure is increasing with increasing Tellurium content, the transition to amorphous phase is completed at 15 GPa for $FeSe_{0.5}Te_{0.5}$ sample and 20 GPa for $FeSe_{0.3}Te_{0.7}$ sample. Furthermore, an intermediate $P2_1/m$ monoclinic phase appears at high tellurium content sample of $FeSe_{0.3}Te_{0.7}$ under pressure while no such monoclinic phase was detected in $FeSe_{0.5}Te_{0.5}$ sample under pressure at ambient temperature. It appears that stability of $P2_1/m$ monoclinic phase under pressure is confined to tellurium rich samples at ambient temperature. However, high pressure and low temperature structural studies are needed to examine the observed phase boundaries at low temperatures. Such studies are going to be extremely useful in future to link high pressure structural transformations to variations in superconducting transitions at low temperatures.

**Acknowledgement:**

We acknowledge support from the National Science Foundation (NSF), Division of Materials Research (DMR), under Grant No. DMR – 0703891. Portions of this work were performed at HPCAT (Sector 16), Advanced Photon Source (APS), Argonne National Laboratory. Andrew Stemshorn acknowledges support from the Department of Education Grant No. P200A090143.

**Figure Captions:**

Figure 1: Angle dispersive x-ray diffraction patterns for FeSe$_{0.3}$Te$_{0.7}$ sample and a copper (Cu) pressure marker at various pressures. All spectrums have been collected at a synchrotron source using a x-ray wavelength λ = 0.3875 Å. (a) Sample in the P4/nmm tetragonal phase at 0.53 GPa, (b) Sample at 12.2 GPa in the monoclinic P2$_1$/m phase along with a weak amorphous component marked by an "asterisk", (c) sample in pure amorphous phase showing three broad peaks denoted by an "asterisk", and (d) sample in amorphous phase during decompression at 5.45 GPa. The diffraction peaks from Cu pressure marker are also indicated and "g" represents weak peaks from spring steel gasket.

Figure 2: The measured variation of the axial ratio (a/b) for the tetragonal and monoclinic phases of FeSe$_{0.3}$Te$_{0.7}$ sample to 20 GPa. The axial ratio (a/b) is unity for the tetragonal phase below 6.4 GPa and increases above this pressure indicating gradual distortion of the tetragonal unit cell.

Figure 3: The variation in the interplanar spacing for the broad band that is observed in the amorphous phase for FeSe$_{0.3}$Te$_{0.7}$ sample between 1 GPa and 30 GPa. The compression and decompression data show a large hysteresis associated with the amorphous phase transition

Figure 4: The measured equation of state (P-V data) for the tetragonal and monoclinic phases for FeSe$_{0.3}$Te$_{0.7}$ sample to 20 GPa at ambient temperature. The measured data points are denoted by open symbols while the dashed curve is a combined fit for both phases to the Birch Murnaghan equation of state with parameters that are described in the text.

.

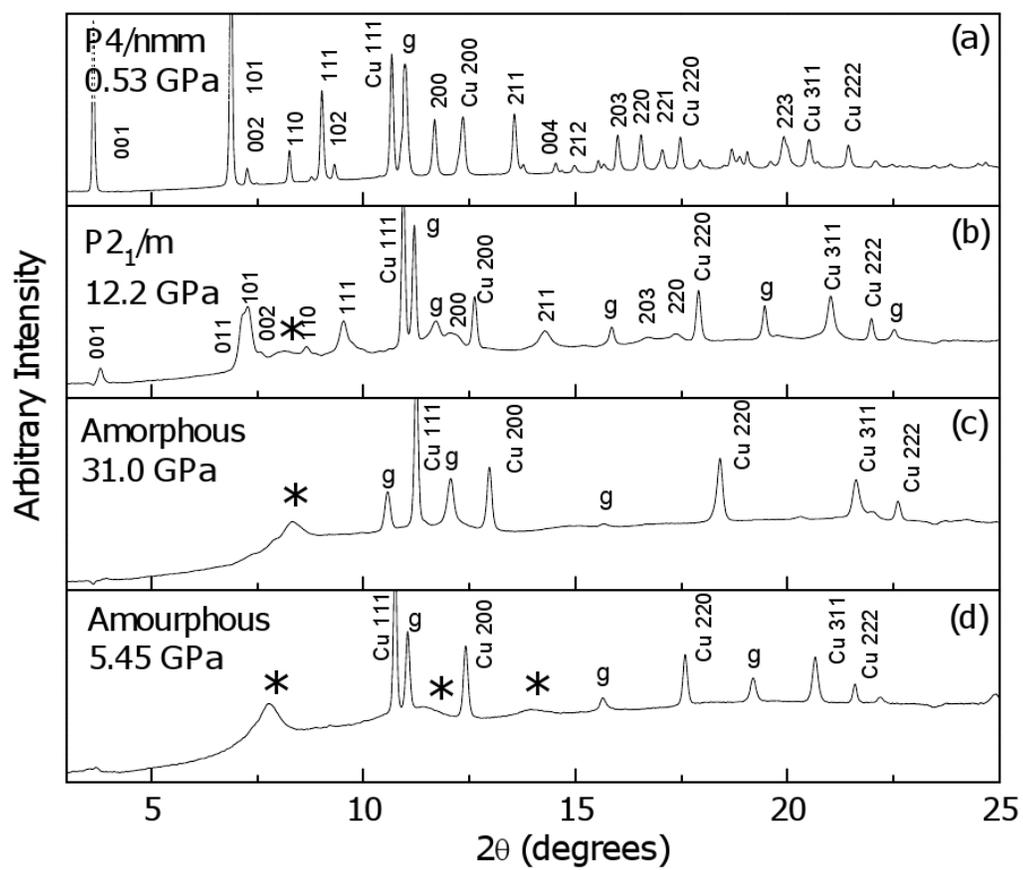

**Figure 1**

10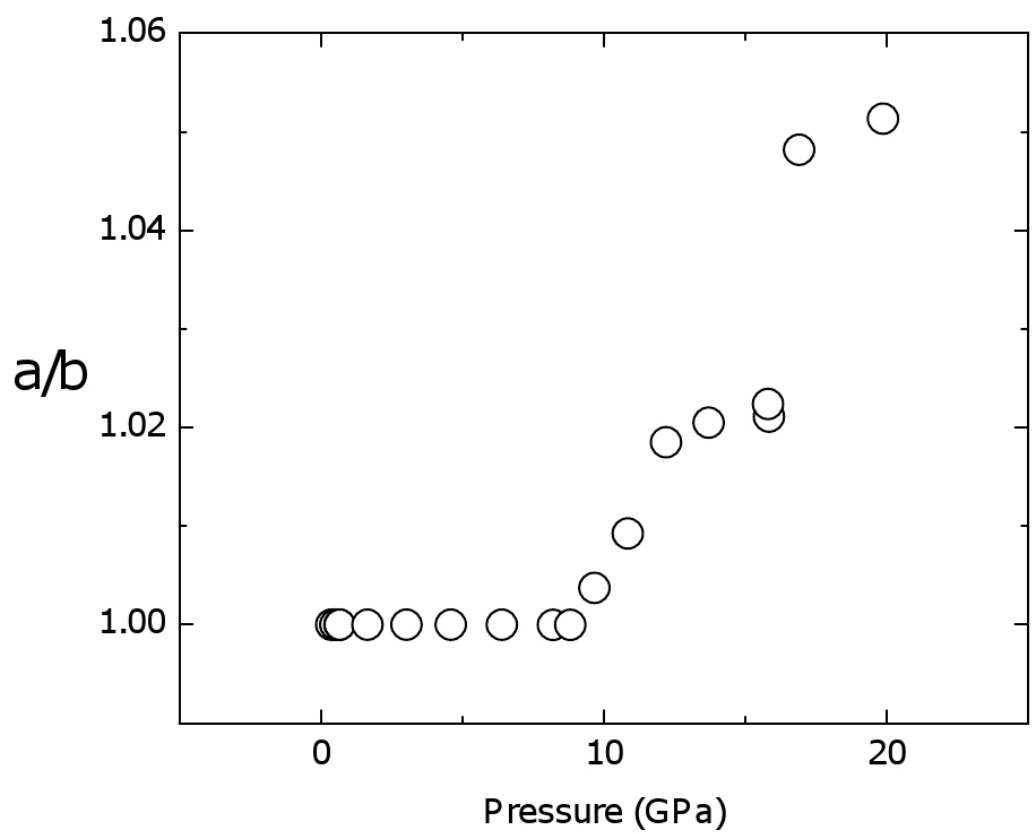

**Figure 2**

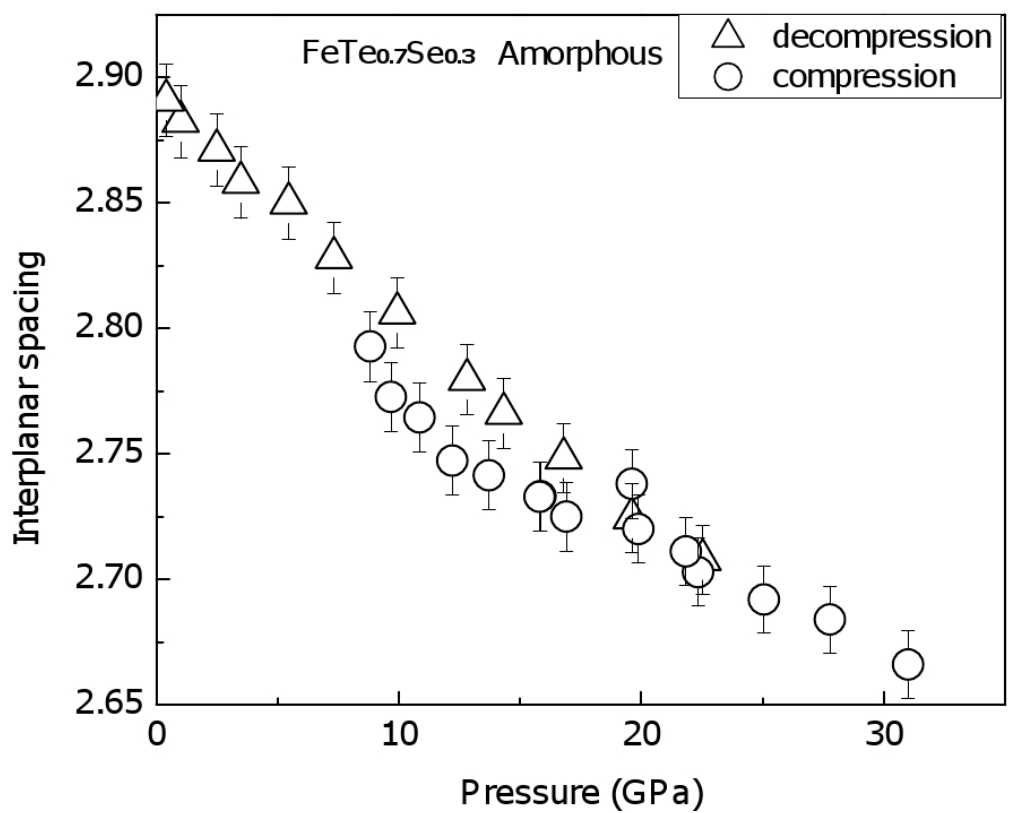

**Figure 3**

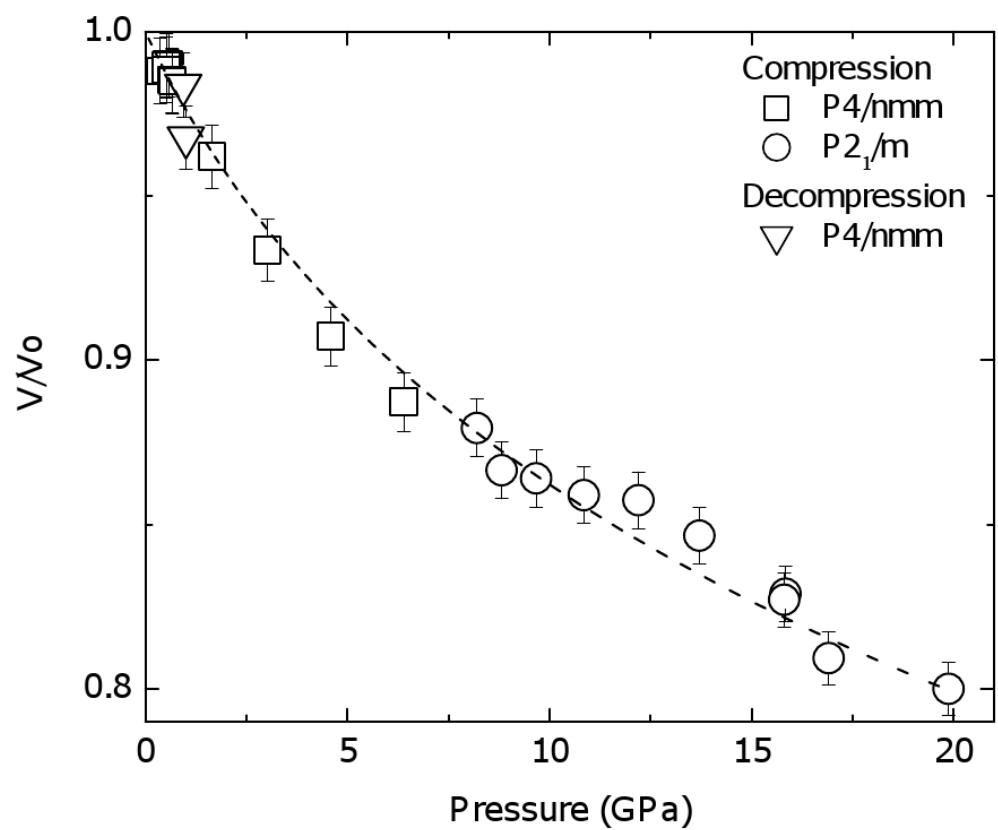

**Figure 4**